\documentclass[11pt]{article}

\usepackage[utf8]{inputenc}
\usepackage[T1]{fontenc}
\usepackage{lmodern}
\usepackage{amsmath,amssymb,amsthm}
\usepackage{mathtools}
\usepackage{booktabs}
\usepackage{graphicx}
\usepackage{geometry}
\usepackage{enumitem}
\usepackage{natbib}
\usepackage{xcolor}
\usepackage[hidelinks]{hyperref}
\theoremstyle{plain}
\newtheorem{proposition}{Proposition}

\theoremstyle{definition}

\theoremstyle{remark}

\newcommand{\E}{\mathbb{E}}

\newcommand{\roas}{\mathrm{ROAS}}

\title{\textbf{Media Measurement and the Assisted Own Goal:\\
Attribution, Marketing-Mix Models,\\
and Individual-Level Incrementality}}

\author{
Tobias B. Konitzer, PhD\thanks{GrowthLoop. Email: \texttt{tobias.konitzer@gmail.com}.}
}

\date{06/21/2026}

\begin{document}
\maketitle

\begin{abstract}
\noindent
We use the \emph{assisted own goal hypothesis} as a lens into media measurement.
A demand-generating (upper-funnel) advertising platform such as a short-video
social network can cause an incremental purchase, yet see that purchase
booked on---and credited to---a downstream trusted marketplace, because
consumers who discover a product on the platform complete the transaction
elsewhere, for example because of distrust of the generating platform as a psychological mechanism. Under attribution-based return-on-ad-spend (ROAS) measurement, the
diverted conversions are invisible to the originating platform. Marketing-mix models (MMMs) do not know which channel to credit with the outcome, and
channel-by-week aggregation denies the audience-level granularity that budget
decisions require. We  develop an incrementality-based measurement
model with two ingredients: \emph{ambient audience-level randomization}---each
activated audience carries its own intent-to-treat (ITT) experiment ---and an \emph{individual-level extension of Predicted
Incrementality by Experimentation (PIE)}, which learns a mapping from
individual features to experiment-identified incremental outcomes. Because ITT
contrasts are computed on channel-complete outcomes, the estimator is
unbiased and the own goal disappears.

\vspace{1em}

\vspace{0.5em}
\end{abstract}

\section{Introduction}
\label{sec:intro}

Media advertising measurement has been plagued by attribution for as long as digital media have existed. Take the canonical example:  The potential customer drives past a  billboard on the famous 101 Highway running through the heart of Silicon Valley again and again, he finally searches for the product, and converts via a search ad (as opposed to organically). How much credit does the ad deserve for the purchase?

A more pertinent example: A user scrolling a feed encounters
a product, forms intent, and completes the purchase on a channel
\emph{other} than the platform where the ad was shown. In particular, consumers
frequently migrate to a large, trusted online marketplace to transact, because
the marketplace offers name recognition, familiar payment, returns, shipping, and buyer
protection that the originating platform's nascent storefront does not [CITE].

These common psychological processes create a measurement pathology. Modern advertisers
disproportionately allocate budget using \emph{attribution-based} ROAS: revenue
is assigned to the advertising touch points that can be observed and matched to a
conversion, typically via pixels, conversion APIs, or last- or multi-touch logic. When the
purchase occurs on a trusted marketplace that does not pass conversion signals from the 
originating platform back to the advertiser  -- and that is true for most inter-marketplace transactions -- the conversion is simply not observed by the
advertiser's attribution system.

The platform that \emph{generated} the demand
emits no return; the marketplace that \emph{captured} or \emph{harvested} the transaction emits records 
the sale. We term the un-credited, demand-generating impression an \emph{assist},
in the football sense: it set up the goal. But because the assist is invisible to
the measurement system in attribution-based systems, it is scored against the originating platform---an
\emph{own goal}. Hence the \emph{assisted own goal hypothesis}: the act of
successfully generating demand can, through the attribution layer, either not increase, or in some cases even reduce the
generating platform's measured performance and therefore its advertising revenue.

The subject of this paper is media measurement.
Advertisers operate three measurement regimes, often simultaneously:
\begin{enumerate}

\item \textbf{attribution}, which assigns credit to observable ad touchpoints;

\item \textbf{marketing-mix modeling} (MMM), which regresses aggregate sales on
channel-level spend; and 

\item \textbf{incrementality measurement}, which identifies
causal effects by randomization. 
\end{enumerate}

The assisted own goal discriminates sharply
    among the three. Attribution cannot see it by construction. MMM, because it
    models total sales, sees the diverted demand in aggregate---but tends to
    re-credit it to the harvesting channel, whose sponsored spend endogenously
    tracks that demand -- akin a multiclolinearity problem in standard regressions. 

Randomized intent-to-treat (ITT) designs
capture the assist correctly (and other measurement pathologies). The constructive half of the paper builds on that
diagnosis: we specify a measurement model in which experimentation is
\emph{ambient} (every activated audience carries its own automatically generated holdout via deterministic hash) and \emph{scaled} to all relevant campaigns.

\paragraph{Contributions.} This paper makes six contributions.
\begin{enumerate}[leftmargin=1.6em,itemsep=2pt]
  \item We formalize the assisted own goal hypothesis in a tractable model of
  consumer journeys with trust-driven channel choice and an explicit
  attribution measurement layer (Section~\ref{sec:model}).
  \item We derive the \emph{attribution wedge}: measured ROAS equals true
  incremental ROAS scaled by an observability factor that is strictly less
  than $1$ whenever trust diverts purchases off-platform
  (Proposition~\ref{prop:wedge}).
  \item We show that ROAS-thresholding budget allocation induces equilibrium
  under-investment in the demand-generating platform, and give conditions
  under which an increase in the platform's \emph{true} ad effectiveness
  lowers its \emph{attributed} revenue---the formal statement of the assisted
  own goal (Proposition~\ref{prop:underinvest} and Section~\ref{sec:owngoal}).
  \item We evaluate the three media-measurement regimes against the own goal:
  attribution misses it by construction; MMM captures the diverted demand only
  in aggregate and can re-credit it to the harvesting channel through
  endogenous sponsored spend; randomized experimentation is invariant to it
  (Section~\ref{sec:mmm}).
  \item We develop an incrementality-based measurement model that scales:
  ambient audience-level ITT randomization via deterministic salted hashing
  (as deployed by GrowthLoop) predicting incrementality from a standing library of experiments. We
  prove the design's immunity to the own goal
  (Section~\ref{sec:itt}, Proposition~\ref{prop:immunity}).
  \item We outline a simulation design under which the attribution bias, the
  induced revenue loss, and the accuracy of the individual-level model can be
  quantified (Section~\ref{sec:sim}).
\end{enumerate}

\section{A Model of Demand Generation, Trust, and Attribution}
\label{sec:model}

\subsection{Players and timing}

There are three types of economic actors:
\begin{itemize}[leftmargin=1.6em,itemsep=2pt]
  \item A \textbf{generator} platform $G$ (e.g., a short-video social network)
  that sells advertising and operates a nascent on-platform storefront.
  \item A \textbf{retailer} marketplace $R$ (e.g., a large trusted online
  retailer) on which consumers can also transact, and which sells its own
  sponsored-placement advertising.
  \item A \textbf{brand} (advertiser) selling a single product at unit margin
  $m>0$, who allocates an advertising budget across $G$ and $R$.
\end{itemize}

\noindent Suppose that: (i) the brand chooses ad spend $s_G$ on $G$ and $s_R$ on
$R$; (ii) advertising generates incremental purchase intent among consumers;
(iii) each consumer with intent chooses a checkout channel; (iv) the attribution
system records conversions it can observe; (v) the brand updates its budget using
measured ROAS. We solve for the brand's measurement-driven allocation and compare
it to the first-best allocation that maximizes true profit.

\subsection{Demand generation}

Let spend $s_G\ge 0$ on the generator produce expected incremental purchase
intent
\begin{equation}
  q(s_G) \;=\; \alpha\, s_G^{\beta}, \qquad \alpha>0,\;\; \beta\in(0,1),
  \label{eq:demand}
\end{equation}
This is a standard demand-generation function, in the sense that increased spend comes with diminishing returns. $\alpha > 0$ is the effectiveness or scale parameter — how much demand a dollar produces overall; better creative or targeting means higher $\alpha$. $\beta \in (0,1)$ is the returns-to-scale exponent, i.e. doubling spend less than doubles the demand created. Each unit of
intent corresponds to one consumer who will purchase the product through
\emph{some} channel. Channel choice is at least partially governed by trust.

\subsection{Trust-driven channel choice}

A consumer with intent generated by $G$ completes the purchase either on $G$'s
on-platform storefront or on the trusted retailer/marketplace $R$. Let $\tau\in[0,1]$ index
the consumer population's \emph{distrust} of on-platform checkout: general reputation, payment
friction, unfamiliarity, weaker returns and buyer protection. Distrust maps directly
into where the transaction is booked:
\begin{equation}
  \Pr\big(\text{booked on } R \mid \text{intent}\big) \;=\; \tau,
  \qquad
  \Pr\big(\text{booked on } G \mid \text{intent}\big) \;=\; 1-\tau.
  \label{eq:divert}
\end{equation}
Note that the total number of incremental purchases is $q(s_G)$ regardless of channel; trust
only governs  \emph{to what extent} the transactions happen on the premises of the generator. This is the heart of the
mechanism: $G$'s advertising is fully effective, but a
share $\tau$ of the resulting sales physically occurs on a competitor's surface.

\subsection{A simple example of demand generation under trust-driven channel choice}
Imagine the following scenario: Acme, a consumer brand, spends a fixed amount on $G$ to generate 1,000 net new potential buyers on TikTik, a social medial network known for sensationalist content and little oversight. Assume the distrust factor  of TikTik is 0.3. which roughly translates to the following: out of 10 buyers  persuaded by ads consumed on TikTik, 3 buyers do not hold enough trust to transact via Acme's storefront on TikTik, and instead transact elsewhere. In our case, 3,000 buyers avoid transaction on TikTik, and instead transact on a trusted marketplace retailer $R$, which for our purposes we shall call Amafon.  

\subsection{The attribution measurement layer}

The brand does not observe true incremental purchases. It observes only what its
attribution framework can match to an ad touchpoint. Let $\phi\in[0,1]$ be the
\emph{recovery rate}: the probability that a purchase completed on $R$ is
nevertheless matched back to the $G$ impression . For purchases completed on $G$'s own storefront the
match rate is $1$. Trusted marketplaces characteristically suppress third-party
signal or share any data of its referrer (in this case $G$), so empirically $\phi$ is close to $0$. The conversions \emph{attributed to} $G$
are
\begin{equation}
  \widehat{q}_G(s_G) \;=\; \underbrace{(1-\tau)\,q(s_G)}_{\text{on-platform, observed}}
  \;+\; \underbrace{\phi\,\tau\,q(s_G)}_{\text{off-platform, partially matched}}
  \;=\; \kappa(\tau,\phi)\, q(s_G),
  \label{eq:attrib}
\end{equation}
where we define the \textbf{observability factor}
\begin{equation}
  \kappa(\tau,\phi) \;\equiv\; (1-\tau) + \phi\,\tau \;=\; 1-\tau(1-\phi)\;\in(0,1].
  \label{eq:kappa}
\end{equation}

In the case of $\phi =0$, the observability factor simply equals $1-\tau$. If trust is at its maximum, all transactions are conducted on the on-platform storefront $G$, and the brand can attribute all conversions to the corret source.

\subsection{True vs.\ measured ROAS}

Define true incremental ROAS and measured (attributed) ROAS as
\begin{align}
  \roas^{\text{true}}_G(s_G) &\;=\; \frac{m\,q(s_G)}{s_G}
      \label{eq:roastrue}\\[4pt]
  \roas^{\text{attr}}_G(s_G) &\;=\; \frac{m\,\widehat{q}_G(s_G)}{s_G}
      \;=\; \kappa(\tau,\phi)\,\roas^{\text{true}}_G(s_G). \label{eq:roasattr}
\end{align}

\begin{proposition}[The attribution wedge]
\label{prop:wedge}
Measured ROAS understates true incremental ROAS by exactly the observability
factor $1-\tau(1-\phi)$ with equality if and only if either there is no distrust ($\tau=0$) or
off-platform purchases are perfectly back-propagated ($\phi=1$). The wedge is strictly increasing in distrust $\tau$ and strictly
decreasing in the recovery rate $\phi$.
\end{proposition}

\subsection{Budget allocation by ROAS thresholding}

The brand allocates spend by a standard rule: invest in a channel up to the point
where its \emph{measured marginal} ROAS falls to a hurdle $\lambda>0$ (e.g. a target ROAS floor). 
Skipping some math, this yields the attribution-driven spend
\begin{equation}
  s_G^{\text{attr}}
  = \left(\frac{m\,\kappa\,\alpha\beta}{\lambda}\right)^{\frac{1}{1-\beta}},
  \label{eq:sattr}
\end{equation}
whereas the first-best spend, which the brand would choose if it observed true
incremental returns, is
\begin{equation}
  s_G^{\text{true}}
  = \left(\frac{m\,\alpha\beta}{\lambda}\right)^{\frac{1}{1-\beta}}.
  \label{eq:strue}
\end{equation}

\begin{proposition}[Equilibrium under-investment]
\label{prop:underinvest}
Under ROAS-thresholding allocation, the demand-generating platform receives
strictly less than first-best spend whenever $\tau>0$ and $\phi<1$:
\[
  \frac{s_G^{\text{attr}}}{s_G^{\text{true}}}
  = \kappa(\tau,\phi)^{\frac{1}{1-\beta}}
  = \big(1-\tau(1-\phi)\big)^{\frac{1}{1-\beta}} \;<\; 1.
\]
This "mismatch" (i.e. decreasing ratio) is worsening with increased distrust $\tau$ and in the diminishing-returns
intensity (i.e.\ increasing in $\beta$), and improving (ie. increasing ratio) in the recovery
rate $\phi$. 
\end{proposition}

\section{The Assisted Own Goal}
\label{sec:owngoal}

The obvious corollary: successful demand generation can lower the
generator's \emph{attributed} revenue, because the demand it creates is booked as
the retailer's performance and triggers a budget reallocation away from $G$.

If we set the price per unit of advertisement to 1, $G$s advertising revenue is exactly the attributed returns in equation 7 above, which is lower than the optimal spend by a factor of $\big(1-\tau(1-\phi)\big)^{\frac{1}{1-\beta}}$ 

\subsection{When effectiveness backfires}

Because G's revenue is simply its attributed returns, anything that lowers measured ROAS lowers $G$'s budget — even if true performance improved. Now suppose $G$ gets better at its job: sharper creative, better targeting, in short a higher $\alpha$. True demand rises one-for-one. But if the very persuasiveness of the ad also lowers average channel trust among the marginal buyers it recruits — if these marginal buyers are disproportionately low-trust consumers with lesser channel affiliation who complete the purchase on the trusted marketplace rather than on-platform, so that $\tau$
 rises with $\alpha$ -- then each gain in effectiveness both grows demand and hides a larger share of it. 
 
 Whenever the measured conversions lost to this unobserved diversion outweigh the measured conversions the extra demand adds, becoming more effective reduces the generator's measured ROAS, its budget, and its revenue. The assist becomes an own goal: G is punished not despite the ad working, but because of it.

\subsection{The reallocation externality}

The diverted purchases do not vanish from the measurement system; they reappear
as $R$'s performance. If $R$ claims a share $\eta$ of the $\tau q$ diverted
conversions through its own last-touch sponsored placements, $R$'s measured
incremental conversions rise by $\eta\,\tau\,q(s_G)$ even though $R$'s own
advertising did not create them. Under the same ROAS-thresholding rule, $R$'s
attributed ROAS is inflated and it attracts budget that, at the margin, is
withdrawn from $G$. The generator thus finances, through its own demand
generation, the measured superiority of its competitor. We summarize the full
mechanism in Table~\ref{tab:mechanism}.

\begin{table}[t]
\centering
\caption{The assisted own goal mechanism, step by step.}
\label{tab:mechanism}
\small
\begin{tabular}{p{0.30\textwidth} p{0.62\textwidth}}
\toprule
\textbf{Stage} & \textbf{What happens} \\
\midrule
1. Demand generation & $G$'s ad creates incremental intent $q(s_G)=\alpha s_G^\beta$. \\
2. Trust-driven diversion & A share $\tau$ of buyers distrust on-platform checkout and transact on $R$. \\
3. Trust-based signal loss & Only $\kappa=1-\tau(1-\phi)$ of true conversions are attributed to $G$. \\
4. Competitor harvest & $R$ books and claims the diverted sales as its own performance. \\
5. Mis-measured ROAS & $\roas^{\text{attr}}_G=\kappa\,\roas^{\text{true}}_G$ understates $G$; $R$'s ROAS is overstated. \\
6. Budget reallocation & ROAS-thresholding shifts spend from $G$ to $R$; $G$'s revenue falls. \\
7. Own goal & Better demand generation can \emph{lower} $G$'s attributed revenue. \\
\bottomrule
\end{tabular}
\end{table}

\section{Media Measurement Regimes and the Own Goal}
\label{sec:mmm}

The assisted own goal is an artifact of last-touch measurement. This section widens the lens across  other popular measurement regimes.

\subsection{Attribution-based measurement}
\label{sec:regime-attr}

Attribution credits conversions to observable, matchable touchpoints---pixels,
conversion APIs, click logs---under a crediting heuristic such as last touch.
We showed at length in the section above that attribution-based measurement regimes are incapable of capturing the conversion assist. It is worth pointing out that the
failure is not the crediting heuristic (i.e. last- vs. multi-touch), but the observability constraint:
while multi-touch or data-driven attribution re-divides credit among \emph{observed}
touchpoints, the own goal removes the conversion from the observable universe. No re-weighting of visible touchpoints can restore credit for this invisible conversion.

\subsection{Marketing-mix models}
\label{sec:regime-mmm}

MMMs regress total sales on channel-level spend across time and geography,
with adstock and saturation transformations and, increasingly, Bayesian priors
\citep{jin2017,chan2017}. Because the dependent variable is \emph{total}
sales---booked anywhere---MMM does not share attribution's blind spot: the
diverted purchases $\tau\,q(s_G)$ remain in the outcome, and in principle the
spend coefficient on $G$ recovers the full demand response.

In practice, three limitations blunt the correction. First, and most directly
tied to our mechanism, the harvester's spend is \emph{endogenous to the
generator's demand}: $R$ prices and sells sponsored placements against the
arriving intent, so any typical regression model regressing total purchase intent on $s_R$ and $S_G$ does not know how much of the
generated demand to attribute to the harvesting channel---the own goal reappears as
simultaneity or multi-colinearity bias rather than signal loss. Second, identification rests on
temporal and geographic variation in spend that national, always-on, or
highly correlated media plans rarely supply. Third, MMM resolves channels by
week or quarter, not by day, so it cannot
drive the thresholding decisions even when its
aggregate reading is correct.

\subsection{Incrementality-based experimentation}
\label{sec:regime-exp}

Randomized designs---geo holdouts, user-level holdouts, ghost ads
\citep{johnson2017}---estimate $q(s_G)$ from the difference in \emph{total}
outcomes between randomized groups, regardless of where transactions are
booked. Such estimates recover $\roas^{\text{true}}_G$ and are, by
construction, invariant to the diversion share $\tau$: a holdout simply buys
less of the product in total, wherever those sales would have occurred. The
regime sees the assist and is immune to the own goal (more below). Its classical
constraint is scale: experiments carry engineering and opportunity costs, and
advertisers will not run experiments for every campaign or audience.

\begin{table}[t]
\centering
\caption{Do the three measurement regimes see the assist?}
\label{tab:regimes}
\small
\begin{tabular}{p{0.20\textwidth} p{0.13\textwidth} p{0.17\textwidth} p{0.38\textwidth}}
\toprule
\textbf{Regime} & \textbf{Sees the assist?} & \textbf{Granularity} & \textbf{Failure mode under diversion} \\
\midrule
Attribution & No & User/campaign, real time & Diverted conversions unobservable; ROAS scaled by $\kappa<1$ (Prop.~\ref{prop:wedge}) \\
Marketing-mix model & Not identified at the channel-level & Channel/week & Harvester's endogenous spend re-credits diverted demand; wide, prior-dominated posteriors \\
Experimentation (ITT) & Yes & Audience/user & None from diversion; binding constraint is cost and coverage \\
\bottomrule
\end{tabular}
\end{table}

\section{An Individual-Level, Intent-to-Treat Measurement Model}
\label{sec:itt}

In this section, we expand on an incrementality-based
measurement model that scales to always-on, audience-level effectiveness measurement and  budget decisions.
It has two ingredients. \emph{Ambient randomization} makes every activated
audience its own experiment, so causal ground truth accumulates as a
by-product of normal marketing operations. \emph{Individual-level prediction}
extends PIE \citep{gordon2023pie} from the campaign grain to the individual
grain, using the accumulated experiments as training data. This set-up assumes that the treatment group of every audience is exported to an ad platform, and each exported audience-member has some probability of eposure on the platform (Intent-to-Treat)

\subsection{Ambient audience-level randomization}
\label{sec:ambient}

Following the design GrowthLoop deploys in production, every audience or
journey activation draws its own control group at the moment it activates.
Assignment is a deterministic hash of the user identifier salted by an
audience-specific key: user $i$ is assigned to control in audience $a$ if and
only if $h(i,a) \bmod 100 < 100\,c_a$, where $c_a$ is the audience's control
percentage. Three properties follow. (i) \emph{Deterministic and stateless}:
the same user always lands in the same bucket for a given audience, with no
assignment bookkeeping. (ii) \emph{Decorrelated}: because each audience uses
its own salt, a user's bucket in audience $a$ is statistically independent of
their bucket in audience $b$, so concurrent experiments overlap only at
random (see \citealp{growthloop_link} for a thorough discussion about SUTVA violations in audience- or campaign-level experimentation). (iii) \emph{Temporally aligned}: each activation draws a fresh split,
so the contrast is causal at any point in the program's life and absorbs the
cumulative effects of earlier campaigns without separate modeling.

\subsection{Intent-to-treat as the estimand}
\label{sec:ittestimand}

Let $Z_i\in\{0,1\}$ denote randomized \emph{assignment} of user $i$ to an
activated audience (eligibility to be targeted), $D_i$ realized exposure, and
$Y_i$ the outcome measured in the brand's first-party transaction
data---\emph{channel-complete} by construction, in the sense that it
aggregates purchases wherever they are booked: on the generator's storefront,
on the marketplace, or offline. The intent-to-treat (ITT) effect
\begin{equation}
  \Delta^{\text{ITT}} \;=\; \E[Y_i \mid Z_i=1] - \E[Y_i \mid Z_i=0]
  \label{eq:itt}
\end{equation}
is identified by randomization alone. Using assignment rather than exposure
avoids conditioning on the ad platform's endogenous delivery decisions (who
saw the ad is algorithmically selected; who was \emph{assigned} is controled by GrowthLoop's randomization procedure), and
matches the advertiser's decision variable: budget buys assignment, not
exposure.

\begin{proposition}[Own-goal immunity of channel-complete ITT]
\label{prop:immunity}
Decompose $Y_i = Y_i^G + Y_i^R$ into purchases booked on the generator's
storefront and on the marketplace. Under randomization of $Z_i$ and
channel-complete outcome measurement, $\Delta^{\text{ITT}}$ identifies the
average incremental purchase effect per assigned user and is invariant to the
diversion share $\tau$, the recovery rate $\phi$, and the marketplace
claim share $\eta$.
\end{proposition}

The proposition states the formal sense in which the own goal is a measurement
artifact: the same assist that is invisible to attribution is fully recoverable by
a channel-complete ITT under a randomized experiment.

\subsection{From campaign-level PIE to individual-level prediction}
\label{sec:pie}

Experiments alone do not cover every campaign, audience, and week. PIE
\citep{gordon2023pie} addresses coverage at the campaign level: using 2,226
Meta RCTs, it trains a model mapping campaign features---including
\emph{post-determined} aggregates such as exposure rates and last-click
conversions, which would be invalid controls in a causal regression but are
valid \emph{predictors} once identification is handled by the
experiments---to experiment-identified incrementality, achieving
out-of-sample $R^2=0.88$ against $R^2=0.19$ for seven-day last-click
attribution. The ambient design of Section~\ref{sec:ambient} supplies exactly
the training resource PIE assumes, but at a finer grain: individual-level
records $(X_i, Z_i, Y_i)$ across a standing library of audience-level
experiments, where $X_i$ collects pre-assignment features (behavioral,
transactional, audience membership) and, for prediction, post-determined
engagement signals.

We therefore extend PIE to the individual level. Define the conditional ITT
function
\begin{equation}
  \delta(x) \;=\; \E[Y_i \mid Z_i=1, X_i=x] - \E[Y_i \mid Z_i=0, X_i=x],
  \label{eq:cate}
\end{equation}
estimable across the experiment library with heterogeneous-treatment-effect
learners. The individual grain nests PIE's campaign grain $\Delta$ by
aggregation: a campaign's predicted incrementality is
$\widehat{\Delta}_a = \E[\hat{\delta}(X_i) \mid i \in a]$.
Beyond nesting, the individual grain adds capabilities the campaign grain
cannot express: ranking users by predicted incrementality when designing the
next audience (e.g. allocating spend toward most persuadable users), transporting predictions to audiences and channels not yet
experimented on, and providing always-on measurement for scaled experiments (see \citealp{growthloop_link_2} for a fragmented discussion about scaling outcomes predictions to always-on measurement). Because fresh ITT experiments continue to accrue ambiently, the model
is periodically re-anchored to experimental ground truth, bounding drift.

\paragraph{Projection onto campaigns without holdouts.}
The projection step is what makes the design scale. The fitted model is a
mapping from observable features to experiment-identified lift. The feature coefficients can now be projected onto any audience with no holdout given that the audience in question has the same features at both individual- and campaign-level as the initial set of experiments.

\subsection{Acquisition advertising and the enumerability constraint}
\label{sec:acquisition}

Ambient user-level randomization requires that the assignment universe be
enumerable at activation. In other words, the universe of buyers that can generate purchases has to be finite and known, in the sense that some individual-level identifier appropriate for ad targeting on the platforms exists. Note this does not have to be NAP (Name, Address, Postal), but could be pseudonymous identifiers like Mobile Ad IDs, or (with limitation) first party cookies that are addressable on the ad platforms.\footnote{An example of this is a flow in which first party cookies are stored within GA4, or assigned by the GA4 Tag, and audiences are onboarded to Google Ads via GA4} 
Acquisition campaigns---lookalike or broad prospecting---violate this: a user
outside the first-party universe can be caused to convert without ever having
been assigned. Neither of the following ex-post fixes works: Assigning converters randomly after
the fact is independent of treatment by construction and dilutes the intent-to-treat effect
toward zero; assigning by observed exposure conditions on the platform's
endogenous delivery  reproduces the attribution bias. Assignment
must precede exposure. 

Note that platform-side lift studies delegate
individual-level randomization to the party that can enumerate at auction
time---though their platform-observed outcomes are not channel-complete, so
the own goal survives inside the lift test itself. Where individuals are
not enumerable at all, the randomization unit can move up to enumerable
aggregates---geographies or time (budget pulses).  ITT on
channel-complete outcomes still would be an unbiased estimator, albeit at a coarser grain. 

Similarly, if the conditional ITT function is the same in both populations (i.e. audience slated for retargeting campaigns and audiences slated for look-alike-based acquisition campaigns), meaning populations differ only in their distribution over features, not in the feature-to-effect mapping, a seed of retargeting audiences can be used as a training data set and inference can be applied over audiences slated for look-alike-based acquisition campaigns with no additional bias.\footnote{This is a strong assumption: it requires common feature support, and it is most strained when the training experiments are retargeting-heavy, since retargeting audiences are selected on the prior engagement that acquisition audiences lack. List-based experimental segments would be a natural bridge.}

In the 
Note that for list-based advertising, where the addressable universe is perfectly contained in the initial audience, no such limits apply.

\section{Simulation Study}
\label{sec:sim}

We close the argument with the simplest simulation the model admits. Consumer
journeys are drawn from the data-generating process of
Section~\ref{sec:model}: spend creates intent ($\beta=0.5$), each intent
converts on the marketplace with probability $\tau$ and on the generator's
storefront with probability $1-\tau$, and the attribution layer matches on-platform purchases
perfectly but recovers diverted purchases only with recovery rate $\varphi$.

\paragraph{The wedge.} For 200{,}000 simulated intents per cell we compute
the ratio of attributed to true conversions across
$\tau\in\{0,.25,.5,.75,.95\}$ and $\varphi\in\{0,0.3\}$. The simulated ratios
(circles in panel a) mimic $\kappa=1-\tau(1-\varphi)$
to three decimal places: at $\tau=0.75$ with no signal recovery, the
generator is credited with exactly one quarter of the conversions it caused.

\paragraph{The backfire.} We let the diversion share rise (read: let the trust decrease) with
effectiveness, $\tau(\alpha)=0.5+0.25\,(\alpha-1)$, hold recovery rate $\varphi$ at $0.1$, and
trace the brand's ROAS-thresholded budget (panel b). First-best spend grows
with $\alpha$ throughout, but attributed revenue peaks at $\alpha\approx 1.7$
and then declines: by $\alpha=2.6$ the generator has lost nearly half of its
peak revenue while being more than twice as effective as at baseline.
Everything to the right of the dotted line is the own-goal region---
where any marginal improvement in effectiveness backfires in terms of attributed ROAS.

\paragraph{Recovery.} Finally we run the ambient experiment of
Section~\ref{sec:itt} at $\tau=0.7$, $\varphi=0.1$: 200{,}000 users per arm,
a 2\% baseline conversion rate on channel-complete outcomes, and a true
incremental effect of 15 conversions per 1{,}000 assigned users. The ITT
contrast estimates $15.4\pm 1.0$---the truth, within sampling error---while
last-touch attribution reports $5.5$, which is $\kappa\times$truth to the
decimal (panel c). Between the two sits the MMM benchmark: aggregating the
same journeys to 600 market-level cells and regressing total sales on
assigned reach and the retailer's sponsored spend---which endogenously tracks
arriving demand---yields $9.7\pm 2.4$, with the shortfall re-credited to $R$:
the coefficient on $R$'s spend is positive even though $R$'s advertising
causes nothing. In other words, MMM sees the diverted demand in aggregate but falsely credits part of
it to the harvester, exactly as Section~\ref{sec:mmm} anticipates.

\begin{figure}[t]
\centering
\includegraphics[width=\textwidth]{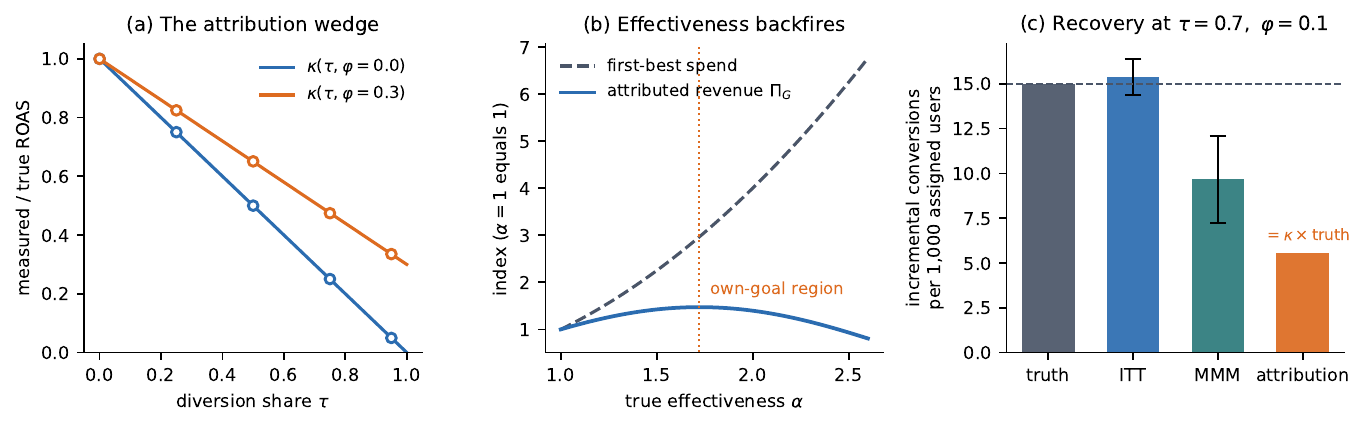}
\caption{The own goal in simulation. (a) Attributed-to-true ROAS ratios from
200{,}000 simulated intents per cell (circles) against the theoretical wedge
$\kappa=1-\tau(1-\varphi)$ (lines). (b) With diversion/distrust rising in
effectiveness, $\tau(\alpha)=0.5+0.25\,(\alpha-1)$ and $\varphi=0.1$; (c) At $\tau=0.7$,
$\varphi=0.1$ across truth, ITT, MMM, and attribution (whiskers: 95\% confidence
interval)}
\label{fig:sim}
\end{figure}

\section{Discussion \& Conclusion}
\label{sec:discussion}
We used the assisted own goal hypothesis as a hook into media measurement.
Under attribution-based ROAS, a platform that successfully generates demand is
penalized when buyers migrate to a trusted marketplace to transact; marketing-mix models see it only in aggregate and can re-credit it to the harvester through endogenous sponsored
spend (multi-co-linearity problem); randomized intent-to-treat designs price it exactly
(Proposition~\ref{prop:immunity}). We proposed a measurement model that makes
the superior regime operational at scale: ambient audience-level randomization
via deterministic salted hashing, as deployed by GrowthLoop, generates a
standing library of ITT experiments, and an individual-level extension of PIE
learns from that library to predict incrementality for users, audiences, and
campaigns and projects it onto audiences and campaigns beyond experimental coverage. 

\paragraph{Implications for advertisers.} Allocating budget on attribution-based
ROAS systematically defunds upper-funnel, demand-generating channels in favor of
lower-funnel channels that harvest existing intent. Advertisers who rely solely on
last- or even multi-touch attribution will mistake the harvesting for the generation.

\paragraph{Implications for measurement practice.} The three regimes can capture true advertising effects with varying degrees of unbiasedness. Attribution-type regimes overstate ROAS of the harvester and understate ROAS of the generator, MMM-type regimes do not know which channel-spend it should allocate observed outcomes/purchases to (akin to the multi-colinearity problem in regressions), ambient audience-level
ITT experiments can capture the dynamics appropriately.

\paragraph{Limitations.} The model is deliberately parsimonious: a single
product, a single generator and retailer, a static one-shot allocation, and a
reduced-form trust parameter. Endogenizing $\tau$ as an equilibrium outcome of
checkout investment, modeling multi-touch attribution windows, and allowing the
retailer to strategically suppress $\phi$ are natural extensions. The measurement model of Section~\ref{sec:itt} adds
its own assumptions: channel-complete first-party outcomes, negligible
cross-experiment interaction (Section~\ref{sec:ambient}), and transportability
of conditional effects to un-experimented audiences; non-changing addressable audiences.


\end{document}